\documentclass[sigconf, authorversion]{acmart}

\usepackage{tabularx}
\usepackage{dcolumn}

\AtBeginDocument{%
  }

\copyrightyear{2025}
\acmYear{2025}
\setcopyright{cc}
\setcctype{by}
\acmConference[IUI '25]{30th International Conference on Intelligent User Interfaces}{March 24--27, 2025}{Cagliari, Italy}
\acmBooktitle{30th International Conference on Intelligent User Interfaces (IUI '25), March 24--27, 2025, Cagliari, Italy}\acmDOI{10.1145/3708359.3712094}
\acmISBN{979-8-4007-1306-4/25/03}

\usepackage{todonotes}
\begin{document}

\title[Evaluating Co-Creativity Between LLMs and Humans in the Generation of Humor]{One Does Not Simply Meme Alone: Evaluating Co-Creativity Between LLMs and Humans in the Generation of Humor}


\author{Zhikun Wu}
\affiliation{%
  \institution{KTH Royal Institute of Technology}
  \city{Stockholm}
  \country{Sweden}}
\email{zhikun@kth.se}

\author{Thomas Weber}
\affiliation{%
  \institution{LMU Munich}
  \city{Munich}
  \country{Germany}}
\email{thomas.weber@ifi.lmu.de}

\author{Florian Müller}
\affiliation{%
  \institution{TU Darmstadt}
  \city{Darmstadt}
  \country{Germany}}
\email{florian.mueller@tu-darmstadt.de}

\renewcommand{\shortauthors}{Wu et al.}

\begin{abstract}
Collaboration has been shown to enhance creativity, leading to more innovative and effective outcomes. While previous research has explored the abilities of Large Language Models (LLMs) to serve as co-creative partners in tasks like writing poetry or creating narratives, the collaborative potential of LLMs in humor-rich and culturally nuanced domains remains an open question. To address this gap, we conducted a user study to explore the potential of LLMs in co-creating memes—a humor-driven and culturally specific form of creative expression. We conducted a user study with three groups of 50 participants each: a human-only group creating memes without AI assistance, a human-AI collaboration group interacting with a state-of-the-art LLM model, and an AI-only group where the LLM autonomously generated memes. We assessed the quality of the generated memes through crowdsourcing, with each meme rated on creativity, humor, and shareability. Our results showed that LLM assistance increased the number of ideas generated and reduced the effort participants felt. However, it did not improve the quality of the memes when humans were collaborated with LLM. Interestingly, memes created entirely by AI performed better than both human-only and human-AI collaborative memes in all areas on average. However, when looking at the top-performing memes, human-created ones were better in humor, while human-AI collaborations stood out in creativity and shareability. These findings highlight the complexities of human-AI collaboration in creative tasks. While AI can boost productivity and create content that appeals to a broad audience, human creativity remains crucial for content that connects on a deeper level. 

\end{abstract}

\ccsdesc[500]{Human-centered computing~Empirical studies in HCI}

\keywords{Human-AI Collaboration, LLM, Co-Creativity, Memes, Humor}
 \begin{teaserfigure}
  \includegraphics[width=\textwidth]{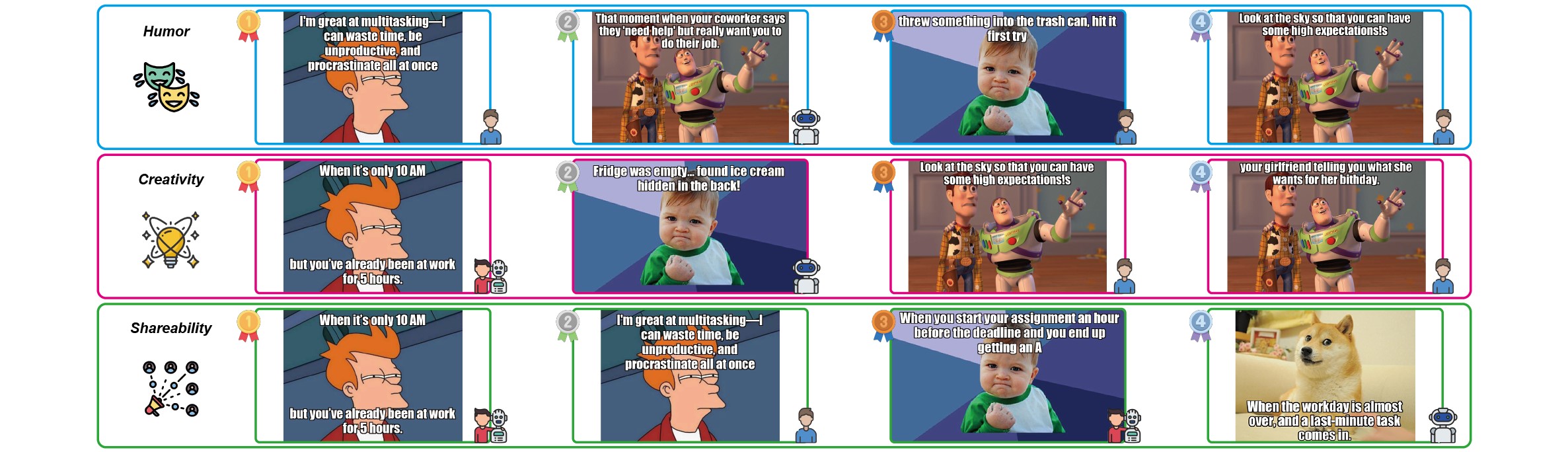}
  \caption{Top 4 Memes Generated by AI, Humans, and Human-AI Collaboration Across Humor, Creativity, and Shareability Metric.}
  \label{fig:teaser}
 \end{teaserfigure}

\received{10 October 2024}
\received[accepted]{12 December 2024}

\maketitle

\section{Introduction}

From co-authoring articles to discussing vacation plans or pair programming, collaborating with other people is a core part of our daily lives in many ways. While we could perform all these activities on our own, the diverse perspectives and problem-solving strategies~\cite{geenSocialMotivation1994}, increased motivation in groups~\cite{geenSocialMotivation1994} and continuous feedback~\cite{wigglesworthWhatRoleCollaboration2012} help to support creative processes~\cite{abraCollaborationCreativeWork1994}. Prior work showed that such collaborative work increases the performance of project teams~\cite{chiocchioEffectsCollaborationPerformance2012}, improves quality~\cite{williamsCollaborativeSoftwareProcess2000}, and, in particular, enhances creativity~\cite{mamykinaCollaborativeCreativity2002} in numerous domains. With the rise of Large Language Models (LLMs), there is a growing trend of replacing human collaboration partners with such systems in typically creative activities in areas such as art~\cite{jiang_2024_haigen,zhu_2018_explainable}, music~\cite{ding_2024_songcomposer, gardner_2023_llark, doh_2023_lpmusiccaps}, and literature~\cite{yuan_2022_wordcraft, wang_2024_weaver, venkatraman_2024_collabstory, hitsuwari_2022_does}. In these areas, LLMs have demonstrated remarkable capabilities in generating content, often matching or even surpassing human performance in tasks like divergent thinking, which involves producing numerous and varied ideas~\cite{hubert_2024_the}.

However, much of the prior work on the creative aspect of LLMs focuses on their outputs as standalone creations, often neglecting their potential as true co-creative partners. In these studies, the researchers presented LLMs with various tasks that typically require creativity and evaluated the result produced by the LLM~\cite{hubert_2024_the, guzik_2023_the,gorenz_2024_how,gmezrodrguez_2023_a}. These assessments focus on attributes like originality and fluency, where LLMs demonstrate strong performance on metrics such as the Torrance Test of Creative Thinking~\cite{hubert_2024_the, guzik_2023_the}. While these works offer important insights into the possibilities of such models for creative tasks, such approaches fail to capture the iterative nature of human co-creativity. Unlike simple task delegation, co-creativity involves iterative co-creation, where humans and AI systems actively refine ideas through dialog and feedback loops, aligning with established frameworks of collaborative creativity\cite{mamykinaCollaborativeCreativity2002,10.1145/1978942.1979214}. Recently, research has started to investigate this co-creative process in the joint creative work of humans and LLMs in a number of domains~\cite{hitsuwari_2022_does,wan_2024_it,liu_2024_how,he_2024_ai} and found that LLMs can contribute novel suggestions, enrich human ideas and enhance the joint creative output~\cite{lu_2024_llm, ali_2023_using}. 

Additionally, one important aspect has not yet been investigated in the area of co-creative collaboration with LLMs: humor. Humor is an interesting area because it is one of the most sophisticated and complex forms of human creativity. Humor strengthens social bonds, addresses difficult topics, and offers new perspectives on everyday situations~\cite{preethamgopalakrishnaadiga_2024_humor}. It relies on surprise, contrast, cultural context, and emotional resonance~\cite{tanaka_2022_learning}. What we think is funny depends on our personal cultural, linguistic and political background~\cite{daviesEthnicHumorWorld1990}. With the rise of online social networks and increasing globalization, parts of this context such as pop culture are aligning across borders, resulting in humor that incorporates elements that are personal and local as well as elements that are globally valid~\cite{shifmanInternetJokesSecret2014}. A well-known example of such humor, that incorporates global and local contexts, are internet memes~\cite{LaughingBorders2016}, often in the form of captioned images. Such memes, as a cultural phenomenon, have emerged as a universal language of the internet and are used to express emotions, convey messages or appropriation and recontextualization  familiar elements~\cite{willmore_2017_internet}. This also made them relevant for research as a means of evaluating creative humor~\cite{vsquez_2021_cats}. However, while prior studies have explored the autonomous generation of memes by LLMs~\cite{wang_2024_memecraft}, there exists no prior work in examining how human-AI collaboration with a multimodal LLM affects the creativity, humor, and shareability of memes.

In this paper, we add to the body of work on human-AI co-creativity by exploring the potential of LLMs as co-creative partners for generating humor.
While 'collaboration' in HCI traditionally entails shared goals and mutual interdependence between multiple human collaborators\cite{mamykinaCollaborativeCreativity2002,10.1145/1978942.1979214}, we follow prior work that defines co-creativity in terms of iterative, dialog-based refinement of ideas~\cite{zhu_2018_explainable,guzdial_2019_an}. In such collaborative systems, the AI can serve as a valuable, but not necessarily equal, partner in the creative process.
Consequently, we investigate how people interact with a ``humor assistant'' in the creation of internet memes and how the availability of such an assistant affects productivity. Additionally, we evaluate how memes generated with such an assistant compare to memes that were created purely by a human and purely by AI in terms of their humor and shareability. 
For this, we conducted two user studies: in the first, we asked participants to generate ideas for memes, either collaboratively, using an LLM, or without any assistance, and rate the experience. 
Our study showed that participants who worked with the LLM assistant generated more ideas during the meme creation process than those who worked independently while perceiving the process as less arduous. 
This first study also yielded 335 images from the human-only group and 307 from the collaborative group which serve as the basis for the subsequent evaluation. For it, we sampled 150 images from each group, as well as 150 fully AI generated images, and asked a second group of participants to rate then in terms of how funny they considered them, how creative, and how likely it would be that they would shared them. This evaluation showed that memes generated entirely by AI surpassed both human and human-AI collaborations in all three dimensions.

These findings indicate that while LLM assistance can significantly boost productivity and reduce perceived effort in creative tasks, it does not necessarily enhance the quality of creative output.
This suggests that AI models, by drawing from vast datasets, are quite adept at producing content that appeals to a wide audience. However, many of the top-rated memes were created with human involvement, suggesting that AI models primarily produce solid but average quality, while human input can help to iterate and curate on it to lift it to a higher level of quality. 






These results emphasize how AI can be tool to quickly and easily produce large volumes of ideas that can often already meed a broad, average appeal. However, it also demonstrates a need for better methods, tools and processes for integrating AI into an iterative creative process where the AI may produce quantity while the humans acts as a curator that selectively pushes towards the AI towards better results. Designing smarter and more human-centered AI systems that can simplify the challenging steps of the creative process while enriching and amplifying humans' unique creative abilities will continue to be a important challenge in the future.

\begin{figure*}[t]
    \centering
    \includegraphics[width=\linewidth]{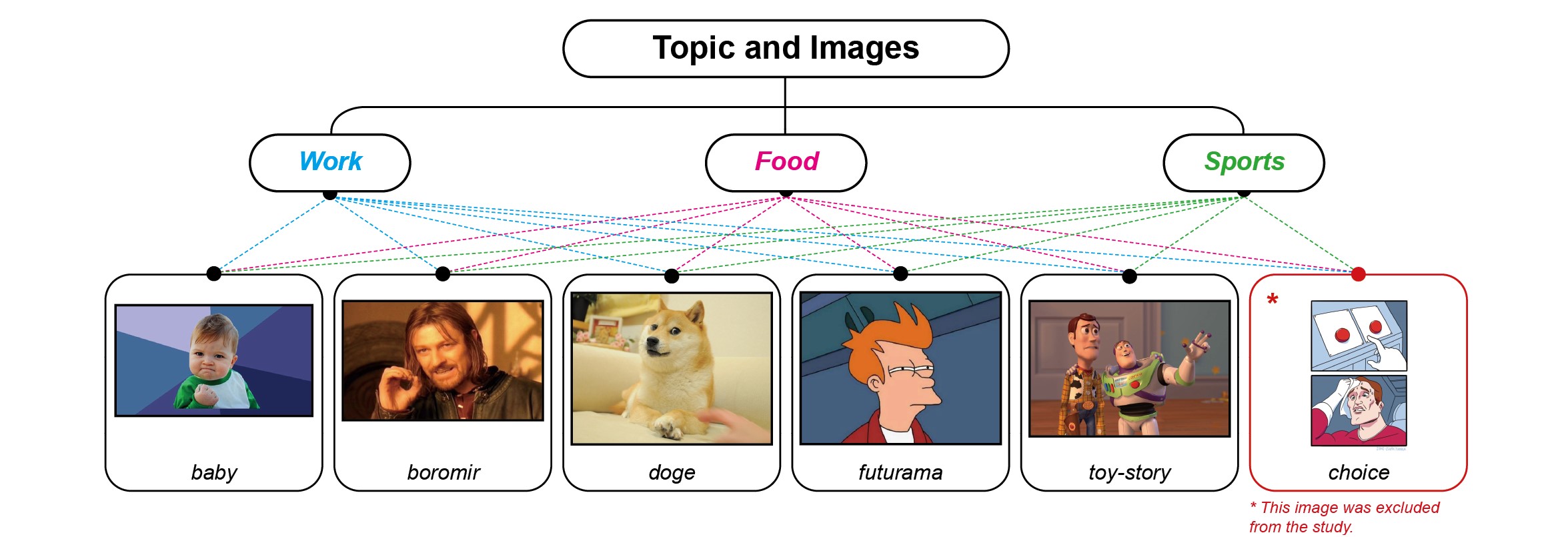}
    \caption{Mapping of Meme Templates to Topics (Work, Food, Sports) in the Study.}
    \label{fig:task}
\end{figure*}
\section{Related Work}

Our research is informed by prior work on human co-creativity, the complexity of humor, human-AI collaboration in creativity, LLMs in creative content generation and the evaluation of creative outputs.

\subsection{Human Co-Creativity and the Complexity of Humor}
Collaborative creativity between humans has long been recognized as a powerful means of enhancing the creative process\cite{10.1145/1978942.1979214}. Co-creativity allows individuals to combine diverse perspectives, skills, and ideas, leading to more innovative and high-quality outcomes~\cite{abraCollaborationCreativeWork1994, mamykinaCollaborativeCreativity2002}.  In group settings, social interactions can stimulate creativity by fostering motivation, providing immediate feedback, and encouraging risk-taking~\cite{geenSocialMotivation1994, wigglesworthWhatRoleCollaboration2012}.

Humor, as a sophisticated and complex form of human creativity, plays a significant role in social bonding and communication~\cite{MARTIN2007113}. Creating humor is particularly challenging because it relies on timing, cultural context, shared knowledge, and the ability to subvert expectations~\cite{doi:10.1177/1088868320961909}. What individuals find humorous is deeply influenced by their personal experiences, cultural backgrounds, and social environments~\cite{daviesEthnicHumorWorld1990}. The complexity of humor makes it a rich area for exploring the dynamics of co-creativity, as collaborators must navigate these nuances to produce content that resonates with others~\cite{KOZBELT2019205}.

\subsection{Human-AI Collaboration in Creativity}

Human-AI collaboration involves humans working alongside AI systems to co-create content by leveraging the strengths of both \cite{florentvinchon_2023_artificial,wu_2021_ai}. In creative fields, this collaboration can enhance human creativity by providing novel ideas and alternative perspectives \cite{guzdial_2019_an}. For instance, in collaborative writing, AI tools generate alternative text suggestions, acting as a "second mind" to stimulate divergent thinking \cite{wan_2024_it}. In design, systems like the Creative Sketching Partner offer visual stimuli to help designers overcome fixation and explore new directions \cite{karimi_2020_creative}.

In the realm of meme creation, however, there is limited research on human-AI collaborative processes. While LLMs can assist in generating meme content, the interplay between human creativity and AI-generated suggestions remains underexplored. Studies in other creative domains suggest that human-AI collaboration can lead to more creative outputs than either humans or AI alone \cite{hitsuwari_2022_does}. Yet, challenges persist, such as managing the AI's lack of contextual sensitivity and ensuring that the collaboration enhances rather than hinders the creative process \cite{jebarezwana_2023_user}.

Moreover, ethical considerations arise in human-AI co-creation, including issues of authorship and bias introduced during AI training \cite{jebarezwana_2023_user, buschek_2024_nine}. Understanding how humans interact with AI systems in creative tasks is crucial for designing tools that effectively support and enhance human creativity.

\subsection{LLMs in Creative Meme Generation}

Large Language Models (LLMs) have demonstrated capabilities in generating human-like text, enabling applications in various creative domains \cite{openai_2023_gpt4}. Recent studies have explored the use of LLMs in autonomous creative content generation, including narratives \cite{yuan_2022_wordcraft,wang_2024_weaver,venkatraman_2024_collabstory}, humor \cite{zhong_2024_lets}, and particularly memes \cite{wang_2024_memecraft}.

In the context of meme generation, MemeCraft \cite{wang_2024_memecraft} utilizes LLMs to produce stance-driven memes with minimal human intervention, showcasing the models' ability to create contextually rich, multimodal content. However, while LLMs can generate humorous and contextually appropriate memes, they often face challenges in capturing nuanced cultural references and emotional subtleties inherent in human creativity \cite{lee_2022_rethinking,esling_2020_creativity}. Studies have indicated that LLMs may produce homogenized content, lacking diversity and originality compared to human-generated content \cite{anderson_2024_homogenization,doshi_2024_generative}.

Despite these limitations, LLMs have been found to outperform humans in certain divergent thinking tasks, exhibiting remarkable originality and fluency \cite{hubert_2024_the, guzik_2023_the}. In humor generation, some LLMs produce jokes rated comparably to human-written humor \cite{gorenz_2024_how,gmezrodrguez_2023_a}, although they may still fall short in capturing the depth of human humor in various contexts \cite{esling_2020_creativity}. These findings highlight both the potential and the challenges of using LLMs for autonomous creative tasks such as meme generation.
\begin{figure*}[t]
    \centering
    \includegraphics[width=\linewidth]{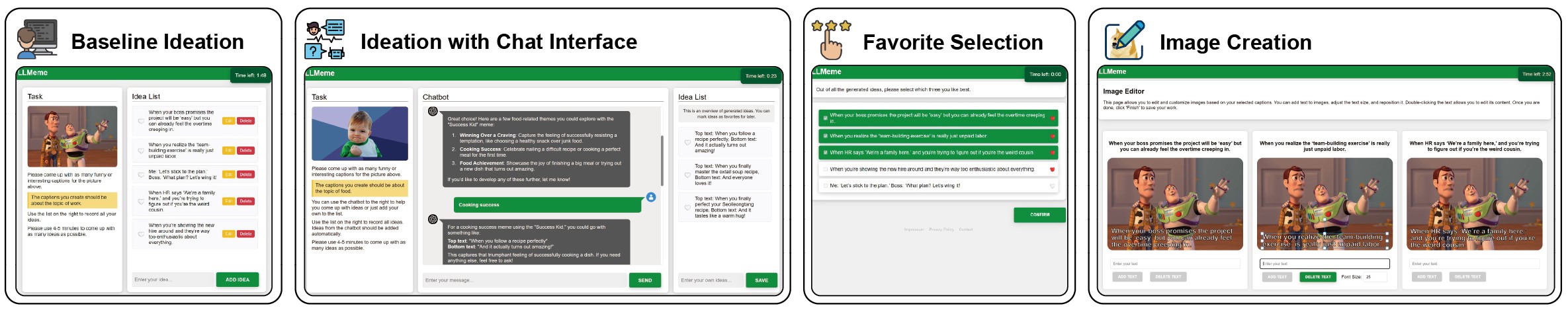}
    \caption{User Interface Overview:Baseline Ideation, Ideation with Chat Interface, Favorite Selection, and Final Image Creation.}
    \label{fig:ui}
\end{figure*}

\subsection{Evaluation Metrics for Creative Outputs}

Evaluating creative outputs, such as memes, involves assessing aspects like creativity, humor, and shareability \cite{vsquez_2021_cats,willmore_2017_internet}. Memes are unique cultural artifacts that blend visual and textual elements to convey messages resonating with diverse audiences \cite{blackmore_2000_the}. The shareability of a meme reflects its potential to be widely circulated, influenced by factors like humor, relatability, and relevance to current cultural topics \cite{molina_2020_what,ling_2021_dissecting}.

Humor is a key driver of engagement and virality in memes, often relying on incongruity and the juxtaposition of unexpected elements \cite{tanaka_2022_learning,malodia_2022_meme}. Memes that effectively utilize humor can facilitate social bonding and amplify sociopolitical discourse \cite{preethamgopalakrishnaadiga_2024_humor}. Creativity, encompassing originality and novelty, is critical in making memes stand out in the vast online content landscape \cite{coscia_2014_average}.

Prior studies have employed both qualitative and quantitative methods to evaluate memes. Qualitative analyses involve content analysis of themes and cultural relevance \cite{tanaka_2022_learning,du_2020_understanding}, while quantitative approaches use machine learning models and sentiment analysis to predict meme virality based on textual and visual attributes \cite{ling_2021_dissecting,chu_2017_predictionasaservice}. However, evaluating creative outputs poses challenges due to the subjective nature of creativity and humor, and the dynamic, context-dependent nature of memes \cite{barnes_2024_topicality}.

\section{Methodology}

To explore the impact of human collaboration with LLMs on creative meme generation, we conducted a between-subject user study with three experimental groups. The following section presents the methodology of the user study.


\subsection{Task}

The participants' task in the study was to generate captions for memes. More specifically, the task consisted of three steps: 

\begin{description}
    \item[Ideation] In the first step, we displayed one of six background images of popular memes(\autoref{fig:task})to the participants and asked them to come up with as many captions as they could within five minutes. We asked participants to focus their ideas on one of three topics: \textit{work}, \textit{food}, and \textit{sports}. The goal was to keep the ideas relatively constrained, for comparability but also not to overwhelm users with having to come up with arbitrary ideas. The interface(\autoref{fig:ui}) displayed a blank meme template as well as the instructions to the user. Users could then enter any ideas for image captions and they were then displayed in a list next to the instructions. Once the user had created ideas, they were also able to mark them as favorite, edit or remove them. For users in the treatment condition that had access to a LLM, this part of the interface also featured a chat interface where they could prompt the LLM(\autoref{fig:ui}). Any responses by the LLM were additionally processed to automatically determine whether the response contained any ideas. If this was the case, they were automatically extracted and added to the idea list. 
    \item[Favorite Selection] Once participants had completed the ideation step, they moved on to an overview of all the ideas they had come up with(\autoref{fig:ui}). From this full list, they had to select their top three ideas. These three ideas were then used in the last step.
    \item[Image Creation] In the last step, we asked our participants to add their ideas as captions to the meme template. The meme editor allowed users to add text to the image in arbitrary chunks(\autoref{fig:ui}). Each chunk could then be positioned and resized, edited or removed. 
\end{description}

Each experimental group used different methods for creating memes, comparing the effects of creativity driven solely by humans, human-AI collaboration, and entirely AI-driven creation. 

The first group (baseline) participants independently generated ideas and created memes without external assistance by an AI tool or otherwise. 
The second group also involved human participants who had to come up with memes but had access to a conversational interface. Through it, they were able to prompt an LLM to support them with generating ideas.
In the third group, ideas were generated fully autonomously by the LLM.

Following the main study, we evaluated the memes generated by the three groups in terms of their funniness, shareability, and creativity using a second online survey.




\begin{figure*}[t]
    \centering
    \includegraphics[width=\linewidth]{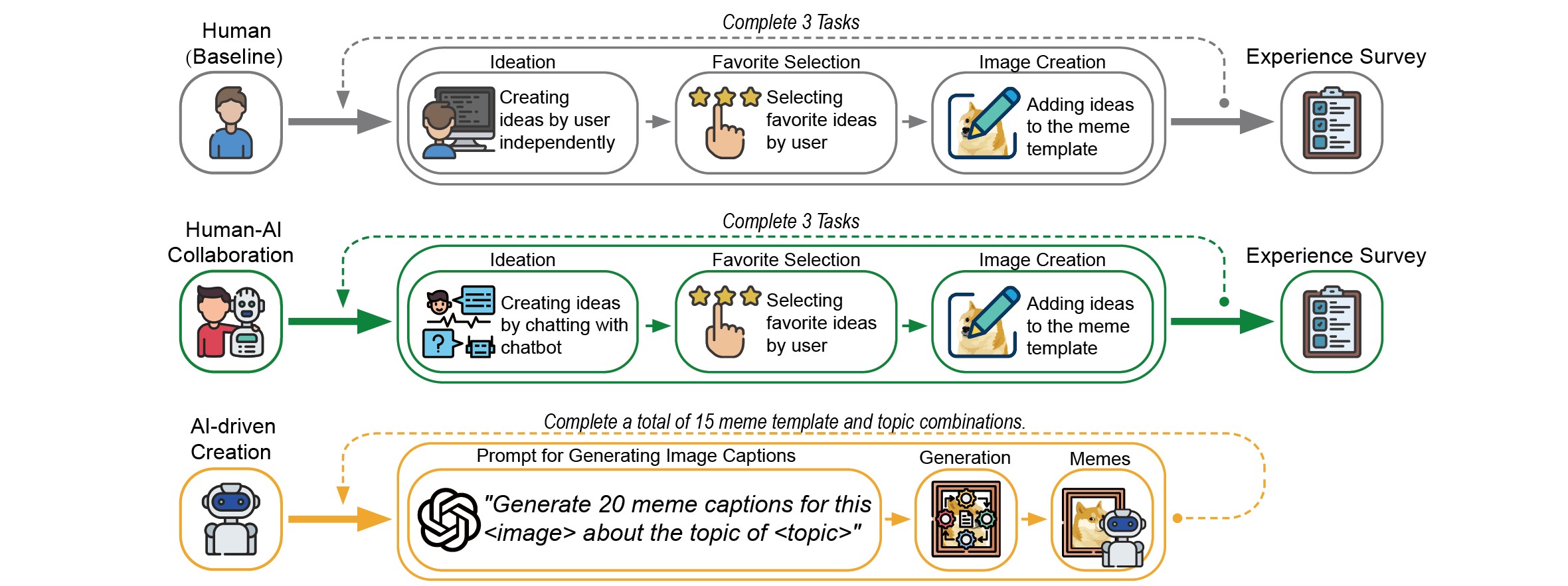}
    \caption{ Meme Generation Workflow: Human (Baseline), Human-AI Collaboration, and AI-Driven Creation. }
    \label{fig:phase1}
\end{figure*}
\begin{figure*}[t]
    \centering
    \includegraphics[width=0.9\linewidth]{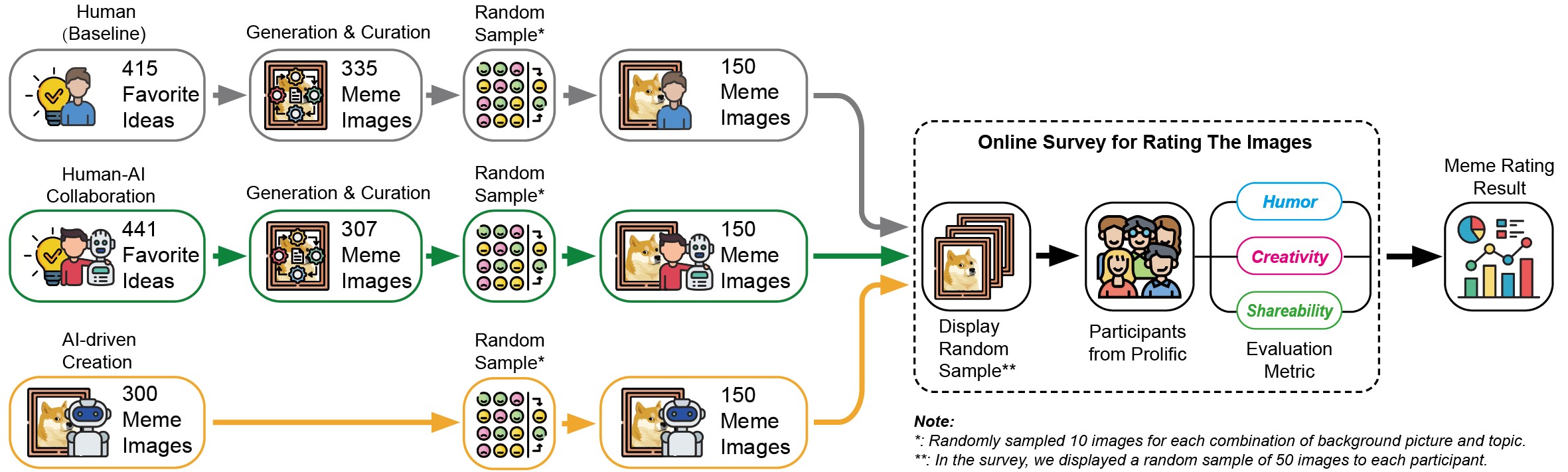}
    \caption{Meme Evaluation Workflow: This diagram illustrates the evaluation process of memes created by humans, human-AI collaboration, and AI-driven approaches.}
    \label{fig:phase2}
\end{figure*}

\subsection{Procedure}

For generating memes, after recording for their informed consent, we asked participants to spend at least four and at most five minutes on coming up with captions using our UI (\autoref{fig:phase1}). Following this ideation phase, they selected their three favorite ideas and could then edit the image to add their idea as captions. After generating and downloading their creation, they moved on to the next image with a next topic. Each participant had to produce memes for three different combinations of images and topics. The permutation of image and topic was selected randomly but in a way that each participant used each image and each topic at most once.
After generating ideas for three different topics and images, participants completed an survey recording feedback on their experience. The overall process of ideation, selecting favorites, editing the images, and completing the survey was scheduled to take no more than 40 minutes. For their work, participants received compensation equivalent to 15 USD.

Following this first phase, we continued with rating the generated ideas (\autoref{fig:phase2}). Since each participant had to selected three favorite ideas for each of the three images they captioned, we gathered 882 ideas marked as favorites across both study conditions with human involvement. Due to technical problems, only 415 for the baseline conditions and all 441 for the collaborative condition were usable.

For each idea, we re-generated the captioned image to ensure consistent placement of the text. We then curated the images, excluding all those were the participants clearly entered a caption not matching the task or where the length of the caption obscured the majority of the image. Since for one of the images, we had to exclude more than two thirds of the images, we decided to fully exclude it from the study. This left us with 335 images from the baseline and 307 images from the collaborative condition.

For rating the quality of these images, we then randomly sampled 10 images for each combination of background picture and topic, which, at five remaining images and three topics, left us with 150 images from the baseline and 150 images from the collaborative condition. 

We then leveraged LLM to create fully AI generated captions for the third study condition. To this end, we prompted the model to generate captions for each combination of image and topic, giving us additional 150 images.

For assessing the subjective quality of these images, we asked a second group of participants to complete an online survey for rating the images. In the survey, we displayed a random sample of 50 images to each participant. For each image, the participated provided feedback along three dimensions: humor, creativity and shareability. These categories were selected based on prior work. We estimated that each rating would take 10--15 seconds, so participants should complete the task in about 10-15 minutes. For their participation, they received compensation equivalent to 10 USD.

\subsection{Prompting}
For conducting the study, we used LLM in two functions: first, as part of the UI where participants could generate ideas with the assistance of a conversational UI. In this interface, participants were free to enter any prompt into the system. However, we set a system prompt to constrain the functionality and output of the system. This system prompt set the context for the LLM, including the fact that the goal of the system was to help users in creating meme ideas, the tone of the interaction to be helpful and polite, and it constrained the system to produce at most three ideas with a single response. Additionally, we always sent the current image to the LLM before any user prompt. The full prompts are available in the supplementary material.

Secondly, we used the LLM to generate image captions for generating the memes for the pure AI condition. For this, we again sent the image first and then instructed the model to \textit{``generate 20 meme captions for this <image> about the topic of <topic>''}, where \textit{<topic>} was one of the three topics and \textit{<image>} was a brief description of the image of no more than 10 words. 
A full list of the generated captions is also part of the supplementary material.

\subsection{Apparatus}

The user interface for the study was implemented using React while any data collection and the interaction with the OpenAI API for GPT-4o was performed by a NodeJS server. All processing of the prompts, randomization of tasks, etc. was performed on the server to ensure the integrity of the data.

Both parts of the study were conducted fully online using our implementation of the meme-creation interface for the first part of the study and a commercial survey platforms for any subsequent surveys. 

\subsection{Data Collection}

While participants created memes, we recorded all ideas they came up with, both text and images, as well as a full log of their interaction with the LLM and its responses on the server. For recording the subjective perception, we used a commercial survey platform. The survey included questions for the participants to self-assess their creativity, as well as the NASA-TLX and general questions about the interface and the ideation process. We used the same platform for rating the generated ideas as well. Demographic data was provided via Prolific, which we used for participant recruiting.

\subsection{Participants}

For this first part of the study, we recruited 124 participants using the online platform Prolific. 26 participants were excluded due to not completing the task. The number of participants was determined after an initial power analysis for the study design. 
Given how the success of humor can be highly dependent on language skill, we selected only participants with good English skills. Additionally, we required participants to have used a LLM interface before at least once, to ensure they would be familiar with the concepts and interactions. This resulted in a diverse participant sample from 30 different countries. Of the participant, 63 indicated to identify as male and 35 as female. The average age was 28.8 years (sd: 8.7).

For the second phase of the study, we simiarly recruited a second set of N=100 participants with the same prerequisites for language skills but knowledge of LLMs was not a requirement. 98 of these completed the task, rating at least 50 images. Participants in this group were equally split between identifying as male and female with an average age of 32.6 (sd: 11.1) and originating in 29 different countries.

\section{Results}

The following section will describe the quantitative findings and their statistical analysis.

\subsection{Meme Creation}


\subsubsection{Idea Generation}

\begin{figure}
    \centering
    \includegraphics[width=0.8\linewidth]{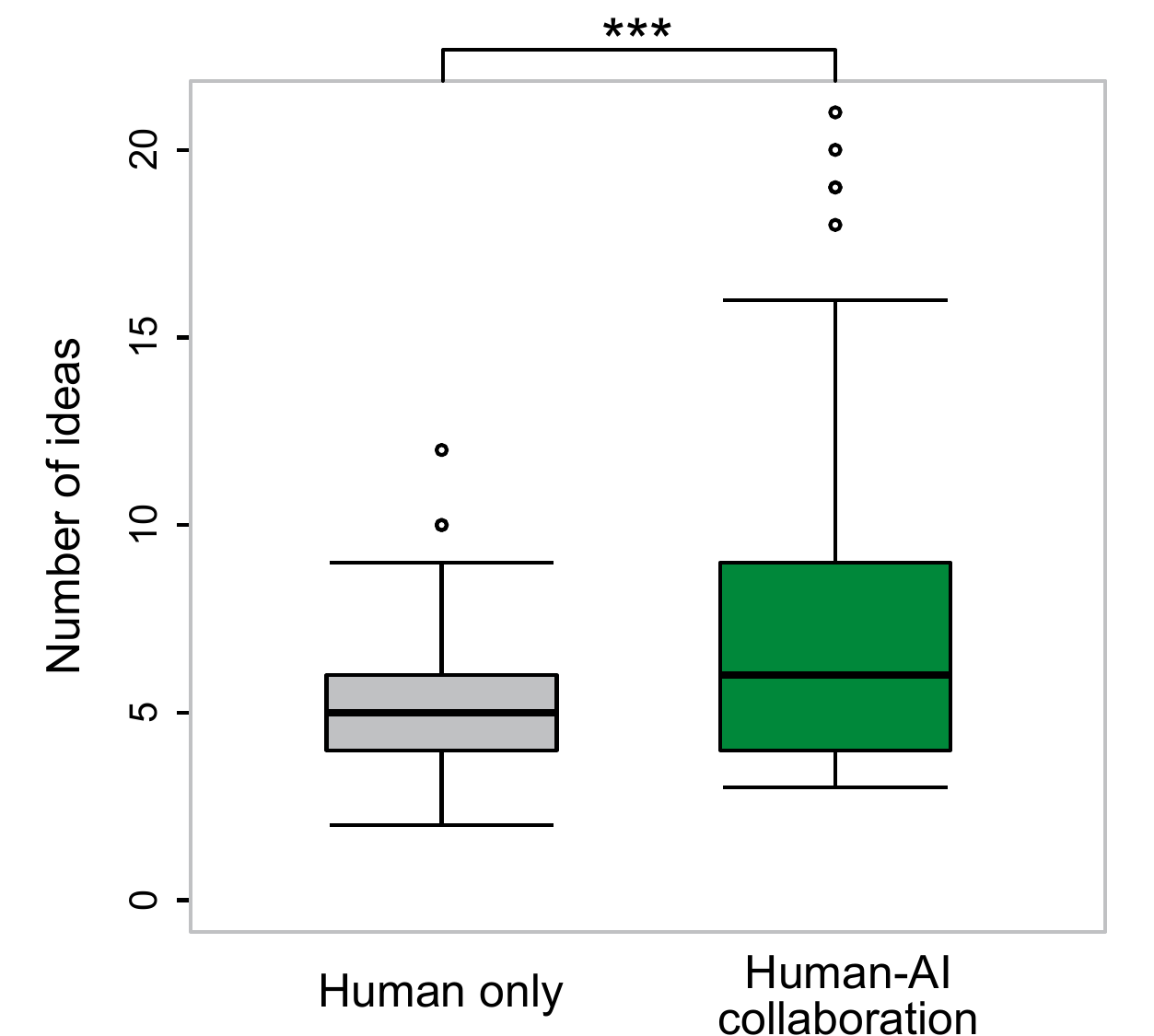}
    \caption{Participants using the LLM were able to produce significantly more ideas than participants who had no external support, according to the Mann-Whitney-U test (***: $p < 0.001$)}
    \label{fig:number-of-ideas}
\end{figure}

\begin{figure}
    \centering
    \includegraphics[width=\linewidth]{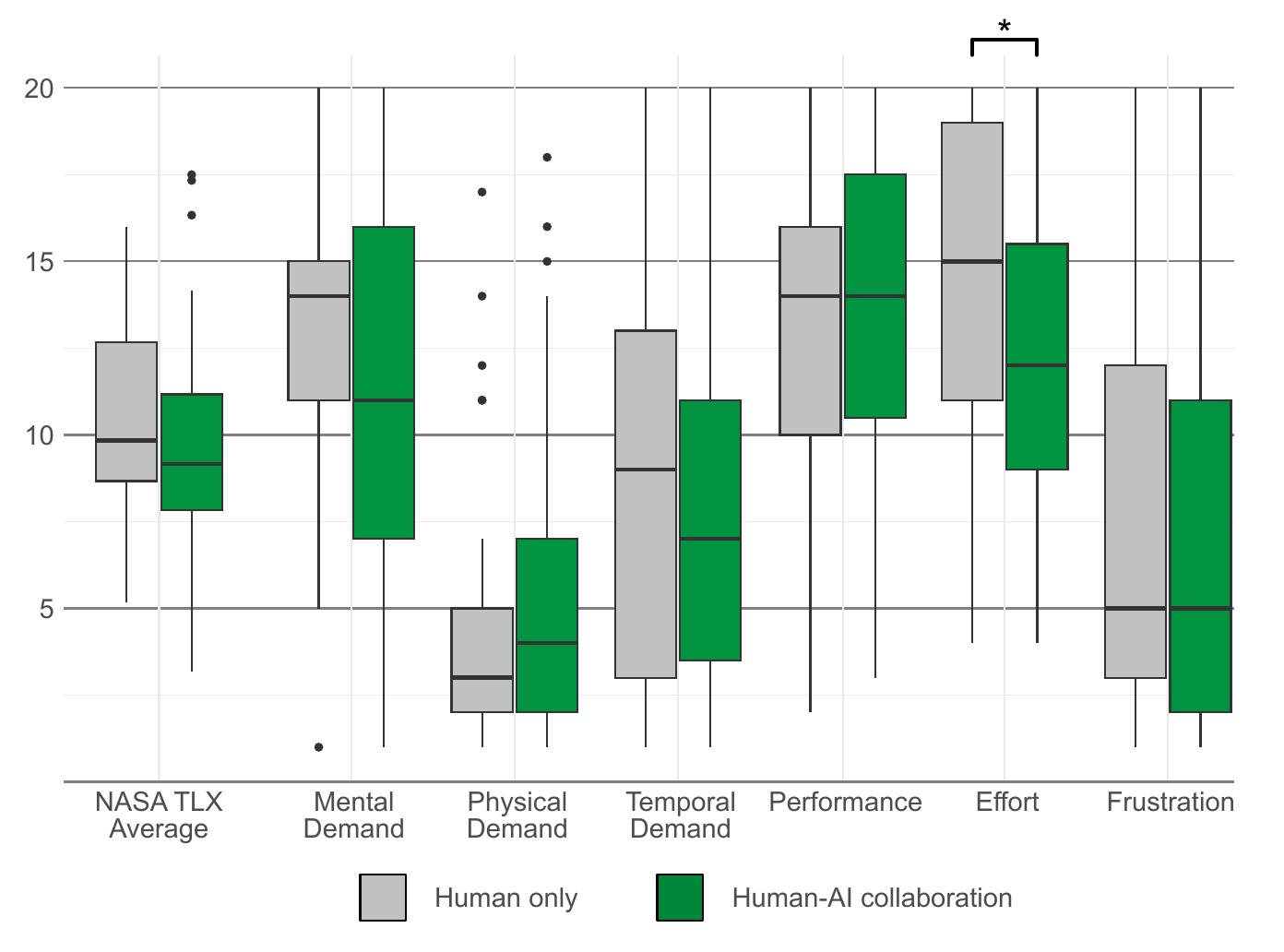}
    \caption{While there were no significant differences in overall workload, the ``Effort'' subscale of the NASA TLX was significantly different according to the Mann-Whitney-U test (*: $p < 0.05$)}
    \label{fig:number-of-ideas}
\end{figure}

\begin{figure*}[t]
    \centering
    \includegraphics[width=0.8\linewidth]{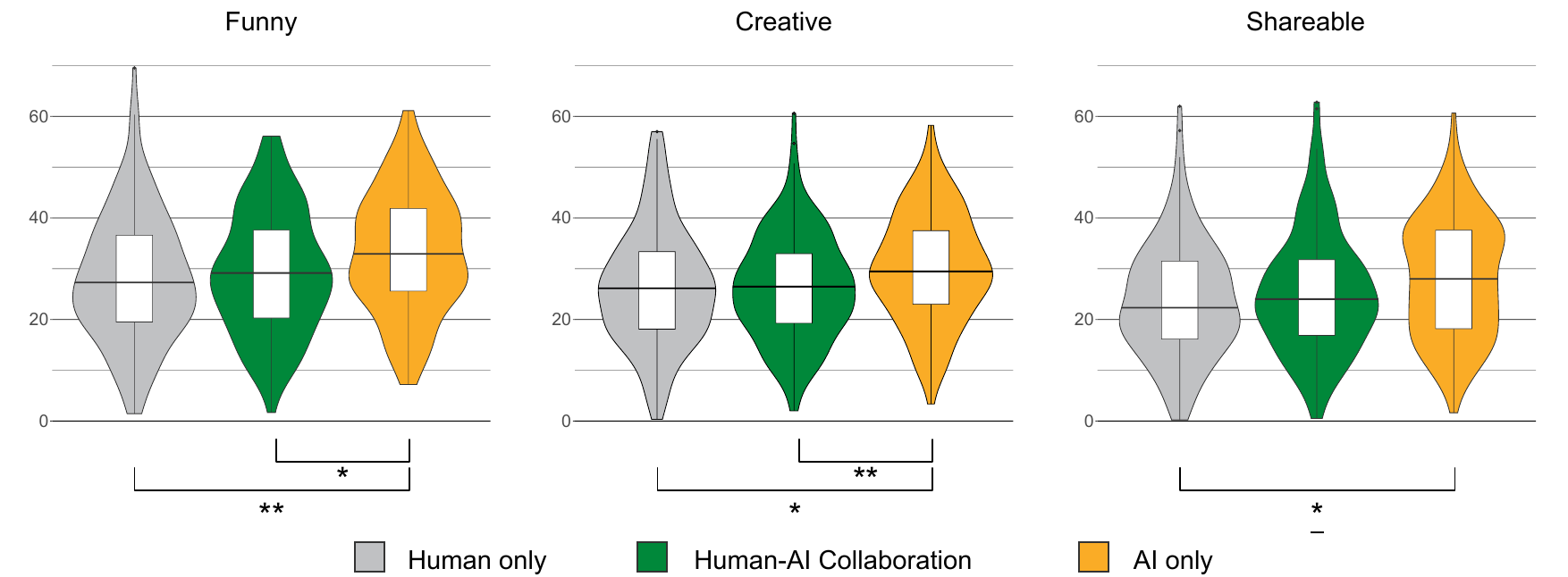}
    \caption{Pairwise comparison of how participants rated the memes with respect to the three scales ``funny'', ``creative'', and ``shareable''. (*: $p < 0.05$, **: $p < 0.01$, **: $p < 0.001$, pairwise t-test/\underline{Mann-Whitney-U test}, Bonferroni adjusted)}
    \label{fig:enter-label}
\end{figure*}

During ideation, participants created an average of 6.1 ideas (sd: 3.2) with one participant managing to come with a total of 21 ideas for one of the images. As seen in \autoref{fig:number-of-ideas}, participants that were able to use the LLM created noticeably more ideas than the participants in the baseline group. To get further insights how the presence of the chat affected the ideation process, we conducted statistical hypothesis tests on the number of ideas per participant. Following a Shapiro-Wilk test to determine non-normality for both the absolute number of ideas ($W=0.811$, $p < 0.001$) and the average number of ideas per participant ($W=0.820$, $p < 0.001$), we used the Mann-Whitney-U test. This test indicated significant differences for the absolute count ($W=12652$, $p < 0.001$) and the average number of ideas ($1519.5$, $p < 0.001$). 

\subsubsection{Workload}

While there are significant differences for the number of created ideas, there is no evidence that this also affected the workload that was required to achieve this result. Statistical analysis of the Raw TLX showed no significant differences (Shapiro-Wilk test: $W = 0.980$, $p = 0.1632$, t-test: $t = -0.955$, $df = 88.811$, $p = 0.342$). We found the same to be true for each of the six TLX subscales, except for the question \textit{``How hard did you have to work to accomplish your level of performance?''}, where participants using the LLM entered significantly lower values (Shapiro-Wilk test: $W = 0.934$, $p < 0.001$, Mann-Whitney-U test: $W = 755$, $p = 0.027$).

\subsubsection{General Feedback}


Similarly, of the general questions about the user's experience  while creating the memes and the ideation process, we received responses that indicated no significant differences except for two questions.

For the first of these was the question whether participants felt that they created a lot of ideas. Results here match the actual idea count, with the LLM-supported users also subjectively noting that they created signiciantly more ideas (Shapiro-Wilk test: $W=0.909$, $p < 0.001$, Mann-Whitney-U test: $W = 1308.5$, $p = 0.043$), although the difference is less stark than with the actual number of ideas.

For the question on perceived ownership \textit{``The generated captions are my ideas''}, participants that did not use the LLM perceived a higher degree of ownership for the generated ideas (Shapiro-Wilk test: $W=0.0.766$, $p < 0.001$, Mann-Whitney-U test: $W = 562$, $p < 0.001$). However, even when using the chat, participants still generally felt ownership for the ideas.

\subsection{Meme Rating}

\begin{table*}[t]
\caption{Results of the statistical analysis of the meme ratings using ANOVA and Kruska-Wallis test (underlined). }
\label{tab:meme-rating-stats}
\begin{tabularx}{\linewidth}{Xrrrrrl}
  \toprule
            & \multicolumn{6}{c}{All topics}                            \\
            \cmidrule(r){2-7}
            & \multicolumn{2}{c}{Shapiro-Wilk test} & \multicolumn{4}{c}{ANOVA / Kruskal-Wallis test} \\
            \cmidrule(r){2-3}
            \cmidrule(r){4-7}
            & W     & p         & df & F/$\chi^2$         & p     &     \\
  \midrule
  Funny     & 0.994 & 0.062     &  2 &             6.971  & 0.001 & **  \\
  Creative  & 0.995 & 0.155     &  2 &             5.793  & 0.003 & **  \\
  Shareable & 0.988 & 0.001     &  2 & \underline{11.761} & 0.003 & **  \\
  \bottomrule
\end{tabularx}
\small
\begin{tabularx}{\linewidth}{lrrrrrlrrrrrrrrrr}
  \toprule
            & \multicolumn{6}{c}{Memes about ``work''}
            & \multicolumn{5}{c}{Memes about ``sports''}
            & \multicolumn{5}{c}{Memes about ``food''} \\
            \cmidrule(r){2-7}
            \cmidrule(r){8-12}
            \cmidrule(r){13-17}
            & \multicolumn{2}{c}{Shapiro-Wilk} & \multicolumn{4}{c}{ANOVA/\underline{KW}}
            & \multicolumn{2}{c}{Shapiro-Wilk} & \multicolumn{3}{c}{ANOVA/\underline{KW}}
            & \multicolumn{2}{c}{Shapiro-Wilk} & \multicolumn{3}{c}{ANOVA/\underline{KW}} \\
            \cmidrule(r){2-3}
            \cmidrule(r){4-7}
            \cmidrule(r){8-9}
            \cmidrule(r){10-12}
            \cmidrule(r){13-14}
            \cmidrule(r){15-17}
            & W     & p         & df & F/$\chi^2$         & p     &
            & W     & p         & df & F/$\chi^2$         & p
            & W     & p         & df & F/$\chi^2$         & p    \\
  \midrule
  Funny     & 0.984 & 0.075     &  2 &             7.1    & 0.001 & **  &
              0.992 & 0.592     &  2 &             2.036  & 0.134 &
              0.977 & 0.014     &  2 & \underline{ 3.455} & 0.177 \\
  
  Creative  & 0.981 & 0.034     &  2 & \underline{11.22 } & 0.004 & **  &
              0.991 & 0.435     &  2 &             2.095  & 0.127 &
              0.985 & 0.095     &  2 &             0.608  & 0.546 \\
  
  Shareable & 0.981 & 0.033     &  2 & \underline{ 8.470} & 0.014 & *   &
              0.983 & 0.063     &  2 &             1.712  & 0.184 &
              0.977 & 0.013     &  2 & \underline{ 4.701} & 0.095 \\  
  \bottomrule
\end{tabularx}
\end{table*}

In the second phase of our experiments, we had a group of people rate the memes according to three criteria: how funny they thought they were, how creative they considered them, and how likely it was that they would share them. 
Along with the memes from the two conditions before, we had an additional condition with memes created exclusively using AI with no human input. 

Considering the fact that the data from the ``funny'' and ``creative'' scale were likely normally distributed (Shapiro-Wilk test: $W = 0.994$, $p = 0.062$ and $W = 0.995$, $p = 0.155$ respectively), we used the ANOVA and pairwise t-tests, Bonferroni adjusted, to compare these two. The third scale, ``shareability'' was likely not not normally distributed (Shapiro-Wilk test: $W = 0.988$, $p < 0.001$), we used the Kruskal-Wallis test instead as well as pairwise Mann-Whitney-U tests, also Bonferroni adjusted.

According to these tests, each condition showed significant differences, as shown in~\autoref{tab:meme-rating-stats}. The pairwise comparison highlighted that it were consistently the memes generated by the LLM alone that were rated more positive than those where created with human involvement. The only exception to this was ``shareability'' where the comparison between the cooperative and pure AI creation was not significant. Memes created by humans with the help of the LLM were not rated significantly different than those from the baseline, i.e. without AI, for any of the three dimensions.  

To ensure any unintended side-effects by of the image or topic selection, we performed the same statistical analysis to determine whether the image or the topic had any notable influence on the rating. This showed that the ratings across the images are relatively consistent, with only one pairing of images showing significantly different ratings for the question how funny the memes were perceived. The topic, on the other hand, seems to have had an impact on the rating of the memes, since work related memes were consistently rated significantly more funny, creative, and shareable.

We therefore analyzed the three scales from the survey again for each topic individually~(see \autoref{tab:meme-rating-stats}), which demonstrated that the previously found significant differences do not come evenly from across all data but seem to stem primarily from the memes about the topic of ``work''.  

\section{Discussion}

The results of our study provide some early insights into how the availability of LLM support influences people's creative process. Further, we investigated how the output of human-LLM co-creation is viewed, compare to purely LLM-generated content. In the following, we discuss our results with regard to the research questions.

\subsection{LLM support increases content output without increasing effort but might diminish the feeling of ownership}
In our study, participants who worked with the LLM assistant came up with significantly more ideas when creating memes compared to those participants working alone. Interestingly, even though they generated more ideas, they did not feel like the task was more demanding. The NASA-TLX results showed that the overall workload was not much different between the two groups, but participants in the LLM-assisted group did report that they had to put in less effort.

These results suggest that using the LLM can make the creative process of crafting humorous content more efficient by helping people generate more ideas without feeling overwhelmed. The AI assistant seemed to help them explore more options without putting in extra effort. This is consistent with previous studies, which suggest that AI tools can support users in generating more creative ideas by reducing obstacles associated with brainstorming and creative development\cite{karimi_2020_creative,guzdial_2019_an}.

Additionally, participants who used the LLM reported a slightly reduced sense of ownership over their creations, indicating that AI assistance might affect the user's connection to their work. However, participants generally still felt that they owned the idea. We attribute this finding to the fact that, in our online testing environment, AI assistance mainly contributed during the idea generation stage, whereas in the stages of idea screening and creating the meme image, no AI assistance or suggestions were provided. The final decision was always left up to the participants. Since feeling a sense of ownership and personal investment plays a significant role in creative motivation and satisfaction\cite{amabile_1983_the}, it is essential to think about how we can balance the involvement of AI in the creative process. 

\subsection{The increased productivity of human-AI teams does not lead to better results - just to more results}

The participants in our study developed more ideas in collaboration with the AI than when they worked alone. However, this did not translate into higher quality in the memes selected by our participants - which always happened without LLM support - in terms of the metrics we collected. This raises questions about the link between quantity and quality in human-AI collaboration. Although coming up with more ideas could increase the chances of producing something high-quality, our study did not find a significant difference. 

In relation to our findings, several prior studies suggest similar outcomes in the context of human-AI co-creativity. For example, Wan et al. \cite{wan_2024_it} found that participants who used an LLM during prewriting activities produced more creative ideas. However, these ideas were not significantly better in quality compared to those created by participants working alone. Similarly, Rezwana and Maher\cite{jebarezwana_2023_user}pointed out ethical and practical challenges in human-AI co-creativity. They noted that while AI can help generate more content, it does not always improve the depth or quality of the work because it struggles with understanding the context and subtleties of creativity. These studies align with our results, showing that while human-AI collaboration can increase productivity and the number of ideas, it does not always lead to higher-quality output. 

From this, we conclude that the use of AI support in the context of the metrics investigated in this study does not lead to better results in terms of humor. Conversely, we can also assume that users with AI support achieve a consistent result faster and with more variations, without this representing an additional mental burden for the users.

\subsection{LLMs appeal to a broad taste in humor, but humans can be wittier still}

Over all assessed metrics, we found that memes created solely by AI performed better than memes created solely by humans or in collaboration between humans and AI. We attribute this initially quite surprising result to the following: The LLM used was trained on large data sets with many cultural references and different types of humor. During the training process, the LLM was most likely to learn the types of humor that it saw most frequently during the training process, i.e. those that resonated best with the crowd. As a result, such LLMs are good at creating content that appeals to a wide audience. The AI picks up on general trends in its training data, which helps it to produce content that most people find appealing. On the other hand, human-created content tends to draw from personal experiences and specific cultural backgrounds\cite{bellaiche_2023_humans}. This broad appeal of AI-generated memes, however, may also be influenced by the composition of the evaluators. The people in the online study, as well as those who evaluated the memes, were random users from a crowdsourcing platform. While this approach brings in a variety of perspectives, it might not capture the subtle differences in humor appreciation across specific demographic groups. Humor is personal and influenced by factors like life experiences, cultural background, and social norms. Therefore, understanding these subtleties would likely require a more targeted evaluation across specific audience segments, which could provide deeper insights into how different groups perceive humor. Even when people work with AI, they often rely on their own experiences when choosing ideas, which creates a similar challenge. It is hard to compete with the AI’s ability to cater to popular tastes.

This interpretation is further supported by the analysis of the top-performing memes (\autoref{fig:teaser}). The funniest memes were mainly created by humans, while those rated highest for creativity and shareability were the result of human-AI collaborations. Among the top 4 humor memes, human creators claimed most of the top positions. In terms of creativity, humans still took half of the rankings, with the rest coming from human-AI teams. For shareability, human-AI collaborations made up half of the highest-ranked memes.

These results highlight that while AI is effective at creating broadly appealing content on average, the individual human touch resonates most deeply on certain dimensions. The top-ranked human creators likely brought in personal experiences, cultural nuances, or innovative ideas that AI, limited to patterns from existing data, cannot fully replicate. At the same time, human-AI collaboration showed real potential, particularly in creativity and shareability, suggesting that AI can offer new ideas or perspectives that, combined with human creativity, lead to content that’s both original and widely appealing.

\section{Limitations and Future Work}

We are convinced that the presented user study provides valuable insights into the creativity process when generating humorous content together with LLMs and the perceived funniness of AI-generated content compared to human (co-) authored content. However, the design and results of our study imply a number of limitations and directions for future work, which we discuss below.

\subsection{Short-term vs. long-term interaction with the AI}

Our study only investigated short-term interactions with the system in a single session. We did not explore how prolonged use of such an AI support system might affect creativity, satisfaction, or the development of new or improved skills. This short-term focus limits our understanding of how users' creative strategies evolve or how they rely on AI over time, which could lead to a decline in quality of the output. Future studies could therefore give participants the opportunity to use AI tools over a longer period of time. By following changes in creative strategies over time, reliance on AI and sense of ownership of generated content, this could provide a better understand the long-term effects of AI collaboration in creative tasks.

\subsection{Limited collaboration between humans and the LLM}

We found that many participants did not fully utilize the potential of the LLM. Some users limited themselves to fulfilling the minimum requirements without revising their ideas in collaboration with the system. Less than half of the participants interacted with the LLM multiple times, and only six participants had more than eight interactions, possibly affecting the quality of their results. A possible reason for this might be our implementation of the study interface. Although the chatbox interface used during the idea generation phase was familiar and easy to use, it lacked structure to guide the creative process. This open-ended approach resulted in varied interactions depending on participants’ individual backgrounds and ideation strategies. Further reasons could include the setting of unambitious goals for the participants, or even an inherent problem with crowdsourcing platforms for conducting such studies~\cite{10.1145/3313831.3376677}. Future work could integrate more structured prompts or more collaborative tools into AI systems to encourage deeper engagement and iterative idea development.

\subsection{Cultural and Social Influence on Humor and Creativity}

What we find funny is heavily influenced by personal backgrounds such as social and cultural factors. In our study, the participants who contributed content and those who rated content came from a wide range of different cultural and social backgrounds. This approach is likely to have led to a wide range of interpretations and preferences for what is considered humorous or creative, thus impacting the outcomes. However, we did not systematically collect detailed demographic information (e.g., ethnicity or language proficiency), limiting our ability to fully explore how varying cultural backgrounds shape humor perception and meme creation. Future work should gather richer demographic and qualitative data—for instance, through open-ended survey questions or interviews—to capture the nuances in how different cultural, linguistic, or social backgrounds perceive and produce humor. Such qualitative insights could also reveal how participants interpret creativity in the context of meme-making, providing a more holistic picture of why certain ideas resonate (or fail to resonate) with different audiences.

\section{Conclusion}

In this paper, we examined the role of LLMs as co-creators in generating humorous content, focusing on the creation of internet memes. Our findings demonstrated that participants who collaborated with an LLM assistant produced a significantly higher number of ideas without reporting an increase in perceived workload, suggesting improvements in both productivity and efficiency. However, this increase in ideas did not consistently lead to higher-quality content when humans were involved. Memes created through human-AI collaboration were rated about the same as those made by humans alone in terms of humor, creativity, and shareability. Interestingly, memes generated entirely by AI scored better, on average, than both human-only and human-AI collaborative memes. But when we looked at the top-performing memes, human-created content was strongest in humor, while human-AI collaborations excelled in creativity and shareability.

These findings show that human-AI collaboration in creative tasks is complex. While AI can increase productivity and produce content that appeals to a wide audience, human creativity is still key for creating content that connects more deeply in certain areas. Participants working with the LLM reported feeling less ownership over their work, suggesting that integrating AI into the creative process needs to be done carefully to keep users connected to their creations. Also, the short-term nature of our study, the limited use of the AI’s full potential due to the open-ended interface, and the similar backgrounds of our participants suggest that more research is needed to understand the long-term effects of AI assistance on creativity and collaboration.

Looking forward, future studies should explore how long-term use of AI affects creative strategies, satisfaction, and skill development. AI interfaces could be improved by providing more structured guidance and encouraging deeper engagement, helping users make better use of AI’s capabilities while maintaining ownership over their work. By addressing these challenges, we can develop smarter, more human-centered AI systems that not only boost productivity but also enhance human creativity.

\begin{acks}
This work was supported by HumanE-AI-Net under Grant Agreement ID: 952026.
\end{acks}

\bibliographystyle{ACM-Reference-Format}
\bibliography{sample-base}


\end{document}